\documentclass[10pt,aps,prd,twocolumn,nofootinbib]{revtex4-1}
\usepackage{amssymb,amsmath,amsfonts,amsbsy,epsfig,graphicx}
\usepackage[dvipsnames]{xcolor}
\usepackage{braket}
\definecolor{lgray}{gray}{0.35}
\definecolor{cor}{rgb}{0.94, 0.41, 0.35}
\definecolor{cas}{rgb}{0.63, 0.72, 0.83}
\usepackage[colorlinks=true,urlcolor=Gray,linkcolor=lgray,
citecolor=Gray,
             pdfpagelabels=true,hypertexnames=true,naturalnames=false,
             ]{hyperref}
\usepackage[utf8]{inputenc}

\newcommand{\be}{\begin{equation}}
\newcommand{\ee}{\end{equation}}
\newcommand{\bea}{\begin{eqnarray}}
\newcommand{\eea}{\end{eqnarray}}
\newcommand{\nn}{\nonumber}

\newcommand{\x}{{\boldsymbol x}}

\DeclareMathAlphabet\mathbfcal{OMS}{cmsy}{b}{n}

\def\x{{\bf x}}
\def\y{{\bf y}}

\def\k{{\bf k}}

\def\N{{\mathcal N}}
\def\bN{{\mathbfcal N}}

\begin{document}

\title{Unifying attractor and non-attractor models of inflation under a single soft theorem} 

\date{November 28, 2022}

\author{
Rafael Bravo$^{a,b}$ and Gonzalo A. Palma$^{b}$
}

\affiliation{
$^{a}$D\'epartament de Physique Th\'eorique and Centre for Astroparticle Physics (CAP), \mbox{Universit\'e de Gen\`eve}, 24 quai E. Ansermet, CH-1211 Geneva, Switzerland.\\
$^{b}$Grupo de Cosmolog\'ia y Astrof\'isica Te\'orica, Departamento de F\'{i}sica, FCFM, \mbox{Universidad de Chile}, Blanco Encalada 2008, Santiago, Chile
}

\begin{abstract}

We study the generation of local non-Gaussianity in models of canonical single field inflation when their backgrounds are either attractor or non-attractor. We show that the invariance of inflation under space-time diffeomorphisms can be exploited to make powerful statements about the squeezed limit of the primordial bispectrum of curvature perturbations, valid to all orders in slow roll parameters. In particular, by neglecting departures from the adiabatic evolution of long-wavelength modes (for instance, produced in sharp transitions between slow-roll and ultra slow-roll phases), we derive a general expression for the bispectrum's squeezed limit in co-moving coordinates. This result consists in the standard Maldacena's consistency relation (proportional to the spectral index of the power spectrum) plus additional terms containing time derivatives of the power spectrum. In addition, we show that it is always possible to write the perturbed metric in conformal Fermi coordinates, independently of whether the inflationary background is attractor or non-attractor, allowing the computation of the physical primordial bispectrum's squeezed limit as observed by local inertial observers. We find that in the absence of sudden transitions between attractor and non-attractor regimes, observable local non-Gaussianity is generically suppressed. Our results imply that large local non-Gaussianity is not a generic consequence of non-attractor backgrounds. 

\end{abstract}

\maketitle


\section{Introduction}

During inflation \cite{Starobinsky:1980te, Guth:1980zm, Mukhanov:1981xt, Linde:1981mu, Albrecht:1982wi} the universe is approximately a de Sitter spacetime. This fact helps to constrain the expected shape of $n$-point correlation functions of the primordial curvature fluctuations $\zeta$, responsible for the existence of structure in our universe. In particular, the de Sitter dilation isometry severely restricts the momentum dependence of $\zeta$'s $n$-point correlation functions~\cite{Creminelli:2011mw, Creminelli:2012ed,Kundu:2014gxa,Kundu:2015xta}.But the existence of an evolving scalar field $\phi(t)$ breaks the de Sitter isometries with departures parametrized by slow roll parameters such as $\epsilon \equiv - \dot H / H^2$ and $\eta = \dot \epsilon/H \epsilon$ (the first and second slow-roll parameters) where $H$ is the Hubble parameter. In many models of inflation, other parameters, such as the sound speed $c_s$ of primordial curvature fluctuations can play a role in breaking the dS isometries introducing effects of order $(1-c_s^2)/c_s^2$. Here we focus our attention on canonical models of inflation, where the sound speed remains equal to 1.  Thus, some statements based on symmetries are not exact (\emph{e.g.} the power spectrum is scale invariant up to corrections of order $\epsilon$ and $\eta$). However, other statements remain valid to all orders in slow-roll. For example, in single field inflation, the primordial bispectrum's squeezed limit (the amplitude of $\zeta$'s 3-point function in momentum space) respects  Maldacena's consistency relation~\cite{Maldacena:2002vr}

\be
\lim_{k_3 \to 0} B_\zeta (\k_1 , \k_2 , \k_3) = -  \big[ n_s (k_1) - 1 \big] P_{\zeta} (k_3 )  P_{\zeta} (k_1) , \label{maldacena-relation}
\ee
where $P_{\zeta} (k)$ and $n_s(k) - 1 \equiv d \ln k^3 P(k) / d \ln k$ are the power spectrum and its spectral index. This result is valid to all orders in slow-roll parameters as long as the background is attractor~\cite{Creminelli:2004yq, Cheung:2007sv}. This can be understood as a consequence of the invariance of Friedmann-Lemaitre-Robertson-Walker (FLRW) spacetimes under a special class of (residual) spatial diffeomorphisms~\cite{Hinterbichler:2012nm, Assassi:2012zq, Creminelli:2012qr, Hinterbichler:2013dpa}. But this understanding is restricted to attractor models, where $\zeta$ becomes constant for wavelengths greater than the Hubble horizon $H^{-1}$~\cite{Weinberg:2003sw}. 

General statements valid for non-attractor models have remained more elusive. In non-attractor models, such as ultra slow-roll inflation~\cite{Tsamis:2003px, Kinney:2005vj,Dimopoulos:2017ged,Morse:2018kda,Firouzjahi:2018vet}, the amplitude of $\zeta$ experiences a rapid growth for wavelengths larger than $H^{-1}$. This property has propelled considerable interest in the study of non-attractor phases during inflation as a way of generating primordial black-holes~\cite{Germani:2017bcs, Hertzberg:2017dkh, Garcia-Bellido:2017mdw, Pattison:2017mbe, Cicoli:2018asa, Biagetti:2018pjj, Byrnes:2018txb, Atal:2019cdz, Atal:2019erb, Motohashi:2019rhu, Ozsoy:2019lyy, Ballesteros:2020qam, Ballesteros:2020sre, Ragavendra:2020sop}. It is well understood that during a non-attractor phase the bispectrum is amplified, leading to a violation of (\ref{maldacena-relation}) which, in the particular case of ultra slow-roll, takes the form~\cite{Namjoo:2012aa, Martin:2012pe, Chen:2013aj, Mooij:2015yka, Romano:2016gop, Bravo:2017wyw, Finelli:2017fml,Esposito:2019jkb}
\be
\lim_{k_3 \to 0} B_\zeta (\k_1 , \k_2 , \k_3) = 6 P_{\zeta} (k_3 )  P_{\zeta} (k_1)  .   \label{maldacena-relation-violated}
\ee
But it has been pointed out that this result does not necessarily stay imprinted in the primordial spectra after the non-attractor phase is over unless the background experiences a sharp transition from the non-attractor phase to the attractor phase~\cite{Cai:2017bxr, Passaglia:2018ixg}. The status of local non-Gaussianity in non-attractor phases of inflation has become a relevant subject with important consequences to our understanding of the early universe~\cite{Huang:2013oya, Sreenath:2014nca, Suyama:2020akr}.

In fact, both (\ref{maldacena-relation}) and (\ref{maldacena-relation-violated}) are expressions strictly valid in co-moving coordinates. As discussed in ~\cite{Tanaka:2011aj} (see also \cite{Pimentel:2013gza}) co-moving coordinates contain spurious couplings between short- and long-wavelength perturbations, altering the derivation of the observable squeezed limit of the bispectrum. To obtain the observable squeezed limit, one may employ a special class of coordinates, the so called Conformal Fermi Coordinates \cite{Pajer:2013ana, Dai:2015rda, Cabass:2016cgp, Bravo:2017gct} (CFC's), allowing the computation of physical quantities observed by inertial observers. As emphasized in \cite{Dai:2015rda}, CFC's remove any diffeomorphism invariance to isolate all locally observable effects, from inflation all the way up to our present epoch. The use of CFC's has been well understood in the case of attractor models of inflation, where the observable primordial bispectrum has been shown to consists of (\ref{maldacena-relation}) corrected by a term $\Delta B =  (n_s - 1) P_{\zeta} (k_3 )  P_{\zeta} (k_1) + \mathcal O (k_3^2/ k_1^2)$, giving us back 
\be
\lim_{k_3 \to 0} B^{\rm obs}_\zeta (\k_1 , \k_2 , \k_3) = 0  + \mathcal O (k_3^2/ k_1^2), \label{B-obs=0}
\ee
where $\mathcal O (k_3^2/ k_1^2)$ stands for projection effects (non-Gaussian contributions due to post-inflationary cosmological evolution). Nevertheless, the use of CFC's to compute the bispectrum resulting from non-attractor phases has remained a challenge. For instance, CFC's were implemented in the particular case of ultra slow-roll in Ref.~\cite{Bravo:2017gct}, where (\ref{B-obs=0}) was indeed recovered. But the computation offered in~\cite{Bravo:2017gct} was limited to the assumption that the universe never abandons the ultra slow-roll phase. Despite of this shortcoming, the result was indicative of a non-trivial cancellation potentially present in more general regimes.

The purpose of this article is to understand the status of the primordial bispectrum's squeezed limit in its three incarnations (\ref{maldacena-relation}), (\ref{maldacena-relation-violated}) and (\ref{B-obs=0}) under a single framework (as we shall see, this framework is achieved by exploiting the diffeomorphism invariance of canoncial single field inflation). In particular, we are interested in the status of (\ref{B-obs=0}) in single field canonical inflation independently of whether the inflationary background is attractor or non-attractor. 

Outstandingly, our approach makes use of time diffeomorphisms (in addition to spatial diffeomorphisms) in order to unify both (\ref{maldacena-relation}) and (\ref{maldacena-relation-violated}) under a single soft theorem (valid as long as the evolution of long wavelength modes remains adiabatic). It is well known that time diffeomorphisms break the co-moving gauge chose to study primordial perturbations~\cite{Hinterbichler:2012nm, Assassi:2012zq, Creminelli:2012qr, Hinterbichler:2013dpa} (see also \cite{Hui:2018cag, Lagos:2019rfc}). However, as we shall point out, one can perform time diffeomorphisms without breaking the choice of co-moving gauge provided that long wavelength perturbations are re-absorbed away into the background.

This article is organized as follows: In Section~\ref{sec:time-diff-FLRW} we analyze the effect of small changes of the time coordinate on the relevant background equations of motion describing single field inflation. In Section~\ref{sec:time-diff-perts} we study how this small change of the time coordinate can be understood as a part of a particular class of space-time diffeomorphism that absorbs long wavelength perturbations in the background. These results are then used in Section~\ref{sec:modulation} to derive a general expression unifying the results  (\ref{maldacena-relation}) and (\ref{maldacena-relation-violated}) into a single soft theorem. Then, in Section~\ref{Sec:CFC} we introduce the so called Conformal Fermi Coordinates and re-examine how long-wavelength modes modulate short modes in such a frame. This allows us to derive some fairly general expressions for the observable primordial bispectrum's squeezed limit in (\ref{sec:obs-bispectrum}). Finally, in Section~\ref{sec:conslusions} we offer some concluding remarks.


\section{Time diffeomorphisms and FLRW backgrounds}  \label{sec:time-diff-FLRW}

We start our analysis by studying the effects of small changes of the time coordinate on background fields. The FLRW metric describing a flat expanding universe is
\be
ds^2 = -  dt^2 + a^2(t) d {\x}^2, \label{background-metric}
\ee
where $a(t)$ is the scale factor, and $d {\x}^2 = \delta_{ij} dx^i dx^j$. In single field inflation the two quantities determining the background configuration are the scalar field $\phi (t)$ and the Hubble parameter $H=\dot a / a$. They satisfy
\bea
\frac{\partial^2  \phi}{\partial t^2} + 3 H \frac{\partial  \phi}{\partial t}  +  V_\phi (\phi)  = 0 ,  \label{background-1}  \\
3 H^2 \!=\! \frac{1}{2} \! \left( \frac{\partial  \phi}{\partial t}  \right)^{\! 2} \!\! + V (\phi), \label{background-2}
\eea
where $V_\phi \equiv \partial V / \partial \phi$. Let us consider the effect of a pure time diffeomorphism of the form:
\be
t \to \bar t = t + \xi^0 (t).  \label{time-diff}
\ee
Given that $\phi$ is a scalar field, this change of coordinates leads to $\phi (t) \to \bar \phi (\bar t) = \phi (t)$. Normally, one would use this relation to write $\bar \phi (t) =\phi (t - \xi^0) = \phi (t) - \dot \phi (t) \xi^0$. Here, instead, we are happy to keep $\bar \phi (\bar t)$ intact, and ask whether it is able to satisfy the same equation of motion as $\phi (t)$, but with $\bar t$ instead of $t$. On  the other hand $a(t)$, which belongs to the spatial part of the metric, does not transform at all under (\ref{time-diff}). That is, in terms of the new time coordinate $\bar t$, the metric (\ref{background-metric}) is given by
\be
ds^2 = -  (1 -2  \dot \xi^0) d \bar t^2 + a^2(t) d {\x}^2,   \label{background-metric-2}
\ee
where the scale factor can be understood as a function of $\bar t$ via $a(t) = a(\bar t - \xi^0)$. Let us now define a new scale factor $\bar a (\bar t)$ as
\be
\bar a (\bar t) = a(t) e^{\alpha} ,  \label{def-new-a}
\ee 
where $\alpha = \alpha (t)$ is a function of time to be determined. So far, $\alpha$ and $\xi^0$ are arbitrary functions. We will fix them in such a way that $\bar \phi (\bar t)$ and $\bar a (\bar t)$ satisfy the same equations of motion~(\ref{background-1}) and ~(\ref{background-2}) obeyed by $\phi(t)$ and $a (t)$, but with $\bar t$ instead of $t$. That is, we demand that $\bar \phi (\bar t)$ and $\bar a(\bar t)$ satisfy:
\bea
\frac{\partial^2 \bar \phi }{\partial \bar t^2} + 3 \bar H \frac{\partial \bar \phi }{\partial \bar t} +   V_{\bar \phi} (\bar \phi) = 0 ,  \label{background-bar-1} \\
3 \bar H^2  = \frac{1}{2} \! \left( \frac{\partial \bar \phi}{\partial \bar t}  \right)^{\!2} \!\! + V (\bar \phi) . \label{background-bar-2}
\eea
Using $\partial t / \partial \bar t = 1 - \dot \xi^0$, we see that to linear order in the diffeomorphism $\bar H (\bar t)= (1 + \dot \alpha / H -  \dot \xi^0) H (t)$. Then, eqs. (\ref{background-bar-1}) and (\ref{background-bar-2}) are satisfied as long as:
\bea
\frac{d}{dt} \left( a^3 \epsilon H \dot \xi^0 \right) = 0 , \label{xi-eq}  \\
3    \dot \alpha   =  (3 - \epsilon ) H \dot \xi^0 . \label{gamma-eq} 
\eea
Thus, there exists a non-trivial time diffeomorphism $t \to \bar t = t + \xi^0$ for which the scalar field $\bar \phi (\bar t) = \phi (t) $ satisfies the same original background equations, but with $\bar a(\bar t)$ instead of $a(t)$. However, notice that (\ref{background-metric}) and (\ref{background-metric-2}) do not share the same form. That is, in (\ref{background-metric}) the coordinate $t$ is cosmic time whereas in (\ref{background-metric-2}) the coordinate $\bar t$ is not. We address this in the following section by including perturbations. 


\section{Time diffeomorphisms and perturbations}  \label{sec:time-diff-perts}

As a next step, let us consider perturbing the same system in two different ways. First, we define perturbations $\zeta$, $\delta \N$, $\bN$ and $\delta \phi$ in such a way that
\bea
ds^2 &=& - e^{2 \delta \N} dt^2 + a^2(t) e^{2 \zeta}(d {\x} + {\bN} dt)^2 , \label{pert-metric-original} \\
\phi &=& \phi (t) + \delta \phi (t , \x) . \label{scalar-pert-def}
\eea
where $(d {\x} + \bN  dt)^2 = \delta_{ij} (dx^i + \N^i dt) (dx^j + \N^j dt)$. In the previous expression $\zeta$ is the spatial curvature perturbation, and $\delta \mathcal{N}$ and $\bN$ are the usual lapse and shift functions. We also consider a second way of perturbing the same metric through perturbations $\bar \zeta$, $\delta \bar \N$, $\bar {\bN}$ and $\delta \bar \phi$ in such a way that 
\bea
ds^2 &=& - e^{2 \delta {\bar \N}} d {\bar t}^2 + \bar a^2(\bar t) e^{2 \bar \zeta}(d {\bar \x} + {\bar {\bN}} d\bar t)^2 ,  \label{pert-metric-new} \\
\phi  &=& \bar  \phi (\bar t) + \delta \bar \phi (\bar t , \bar {\x}) , \label{scalar-pert-def-2}
\eea
where $\bar \phi (\bar t) = \phi (t)$. Here $\bar t$ is the same time coordinate defined in (\ref{time-diff}), and $\bar a (\bar t)$ is the same scale factor defined in (\ref{def-new-a}), with $\xi^0$ and $\alpha$ satisfying~(\ref{xi-eq}) and (\ref{gamma-eq}). We have also introduced the spatial coordinate ${\bar x}^i$ as
\be
\bar \x = e^{\beta /3} \x , \label{space-diff}
\ee
where $\beta = \beta(t)$ is a function of time to be determined [a more standard approach would have consisted in writing ${\bar x}^i = x^i + \xi_s^i$, with $\xi_s^i \equiv \frac{x^i}{3} \beta (t)$ so that $\partial_i  \xi_s^i  = \beta(t)$].  It should be clear that both (\ref{pert-metric-original}) and (\ref{pert-metric-new}) are just different ways of expressing the same metric. However, both the background fields and perturbations differ. This is the key aspect that we exploit in what follows.

We now fix the gauge. We choose to work in co-moving gauge, whereby $\delta \phi$ of Eq.~(\ref{scalar-pert-def}) satisfies $ \delta \phi (t , \x)  = 0$. Given that $\bar \phi (\bar t) = \phi (t)$, this condition implies that $\delta \bar \phi$ in Eq.~(\ref{scalar-pert-def-2}) satisfies $ \delta \bar \phi ( \bar t , \bar \x)  = 0$. Thus, both $\zeta$ and $\bar \zeta$ are comoving curvature perturbations. Is this even possible? As we shall see, (\ref{xi-eq}) and (\ref{gamma-eq}) play an important role in allowing this. Indeed, notice that (\ref{space-diff}) implies that $d \bar \x = e^{\beta /3} d \x + \frac{1}{3} \x \dot \beta e^{\beta /3} dt$,  which in (\ref{pert-metric-new}) leads to
\bea
ds^2 &=& - e^{2 ( \delta {\bar \N} + \dot \xi^0)} dt^2 \nn \\
&& + a^2 (t)  e^{2 ( \bar \zeta + \alpha + \beta / 3)}\left[ d \x  +  {\bar {\bN}} dt + \frac{1}{3} \x \dot \beta   dt  \right]^2 \!\!, \quad \label{metric-step}
\eea
where we also made use of (\ref{time-diff}) and (\ref{def-new-a}). Comparing Eq.~(\ref{metric-step}) with Eq.~(\ref{pert-metric-original}) gives us the following three relations:
\bea
\zeta = \bar \zeta + \alpha + \frac{1}{3} \beta , \label{cond-perts-1} \\
\delta \N =  \delta {\bar \N} + \dot \xi^0 , \label{cond-perts-2}  \\
 \bN = {\bar {\bN}}  + \frac{1}{3} \x \dot \beta  . \label{cond-perts-3}
\eea
Now, let us split $\zeta$, $\delta \N$ and $\bN$ appearing in the left hand side of the previous equations into short- and long-wavelength modes: 
\bea
\zeta (t, \x) &=& \zeta_S (t, \x) + \zeta_L (t), \\
\delta \N (t, \x) &=& \delta \N_S (t, \x) + \delta \N_L (t), \\
\bN (t, \x)  &=& \bN_S (t, \x) + \bN_L (t, \x) .
\eea 
The long-wavelength part $\zeta_L (t)$ corresponds to the zeroth order term of the Taylor expansion $\zeta_L (t, \x) = \zeta_L(t) + \partial_i \zeta_L(t) \x^i + \frac{1}{2} \partial_i \partial_j \zeta_L(t) \x^i \x^j$, and must satisfy the equation of motion for $\zeta$ in the long-wavelength limit:
\be
\frac{d}{dt} \left( \epsilon a^3 \dot \zeta_L  \right) = 0 . \label{motion-long}
\ee
On the other hand, $ \delta \N_L (t)$ and $\bN_L (t, \x)$ must satisfy the constraint equations:
\bea
\delta \N_L  (t) &=&  \frac{1}{H} \dot  \zeta_L  (t),   \label{constraint-1} \\
\bN_L (t, \x) &=& \frac{1}{3} \epsilon \, \x  \, \dot  \zeta_L  (t).  \label{constraint-2}
\eea
Given that we are interested in making statements about the action of $\zeta$ valid up to cubic order, it is enough to express the lapse and shift function linearly with respect to $\zeta$ (see Ref.~\cite{Maldacena:2002vr}). But notice that $\dot \xi^0$, $\alpha$ and $\beta$ in the right hand side of~(\ref{cond-perts-1}) can be adjusted to absorb the long wavelength perturbations $\zeta_L (t)$, $\delta \N_L (t)$ and $\bN_L (t, \x)$ appearing at the left hand side of the same equations. That is, we can impose:
\bea
\dot \xi^0  =  \frac{1}{H} \dot  \zeta_L ,  \label{cond-perts-again2-1} \\
\alpha + \frac{1}{3} \beta  = \zeta_L , \label{cond-perts-again2-2} \\
 \dot \beta  = \epsilon  \dot  \zeta_L . \label{cond-perts-again2-3}
\eea
Let us recall that $\xi^0$ and $\alpha$ are restricted to satisfy (\ref{xi-eq}) and (\ref{gamma-eq}), so it is not obvious that (\ref{cond-perts-again2-1})-(\ref{cond-perts-again2-3}) can hold (even if $\beta$ is a function to be fixed freely). However, $\zeta_L $ respects the equation of motion (\ref{motion-long}) which, thanks to (\ref{cond-perts-again2-1}), coincides with (\ref{xi-eq}). Similarly by taking a time derivative of (\ref{cond-perts-again2-2}), and combining the result with Eqs.~(\ref{cond-perts-again2-1}) and (\ref{cond-perts-again2-3}) we obtain $3 \dot \alpha    = (3 - \epsilon)  H \dot \xi^0 $, which is precisely (\ref{gamma-eq}). 

Thus, we conclude that (\ref{cond-perts-again2-1})-(\ref{cond-perts-again2-3}) [together with (\ref{motion-long})] are consistent with (\ref{xi-eq}) and (\ref{gamma-eq}). A direct consequence of this, is that the remaining spatially dependent functions appearing in (\ref{cond-perts-1})-(\ref{cond-perts-3}) satisfy:
\be
\bar \zeta = \zeta_S   ,  \qquad \delta {\bar \N} = \delta \N_S   ,  \qquad
{\bar {\bN}}  = \bN_S  . \label{identification-zeta-S-zeta-bar}
\ee
This means that the short-wavelength perturbations of the metric (\ref{pert-metric-original}) can be identified as the full perturbations of the metric (\ref{pert-metric-new}). In other words, $\bar \phi (\bar t)$ and $\bar a(\bar t)$ are the background fields felt by $\zeta_S$, $\delta \N_S$ and $\bN_S$, once the long-wavelength modes have become part of the background. Given that the full action describing single-field inflation is invariant under space-time diffeomorphisms, the action for $\zeta$ derived using (\ref{pert-metric-original}) and the action for $\bar \zeta = \zeta_S$ derived using (\ref{pert-metric-new}) have the same form (at least at cubic order). However, the background quantities in the action for $\zeta_S$ contain $\zeta_L$ as part of it.

\subsection{Setting initial conditions} \label{sec:initial-conditions}

Having determined the equations that gives us $\xi^0$, $\alpha$ and $\beta$ in terms of $\zeta_L$, we can proceed to solve them. This requires knowledge of the initial conditions for $\xi^0$, $\alpha$ and $\beta$. From its definition in (\ref{def-new-a}), we may set initial conditions in such a way that $\alpha = 0$ for a given choice of time $t_*$:
\be
\alpha (t_*)  = 0 . \label{alpha-t*}
\ee
This simply ensures that the scale factor $\bar a$ coincides with $a$ at a given time $t_*$. That is:
\be
\bar a(\bar t_*) = a(t_*). \label{a-t*}
\ee 
Given that $a$ cannot be measured directly, we can choose $t_*$ arbitrarily. However, in order to weigh the amount of expansion experienced by short-wavelength modes with $\zeta_L$ absorbed in the background, it is convenient to choose $t_*$ to be a time such that the cutoff wavelength $\lambda_L =  2\pi a(t) / k_L$ separating long- and short-wavelengths is already super-horizon [\emph{i.e.} $\lambda_L(t_*) \gg H^{-1} (t_*)$].  For definiteness, $t_*$ could be taken as the time at which a given short mode of interest crosses the horizon. With this understanding, Eq.~(\ref{alpha-t*}) gives 
\be
\beta (t_*) = 3 \zeta_L (t_*) .
\ee

Finally, because $\xi^0$ corresponds to a reparametrization of time, its form is dictated by the solution of Eq.~(\ref{cond-perts-again2-1}) up to an integration constant which we are free to fix. We examine this in the next discussion.

\subsection{An explicit expression for $\xi^0$}  \label{sec:xi0}

It is possible to derive an explicit expression for the time diffeomorphism $\xi^0$. This can be done by directly integrating $\dot  \xi^0 =  \dot \zeta_L /H$ to obtain an analytical expression in terms of slow-roll parameters to all orders. To do so, we try the ansatz
\be
\xi^0 = C_1 + F \, \dot \zeta_L , \label{new-xi}
\ee
where $C_1$ is a constant of integration and $F = F(t)$ is a function of time to be determined. By taking a derivative of this ansatz, and matching it with $\dot \xi^0 = \dot \zeta_L / H$ we obtain $H \dot F \dot \zeta_L  +  F H  \ddot \zeta_L = \dot \zeta_L$. But recall from (\ref{motion-long}) that $\ddot \zeta_L = - 3 H \dot \zeta_L - H \eta \dot \zeta_L $. Then, $F$ must respect the following equation $H \dot F   - (3 + \eta) F H^2  -1  = 0$. The solution to this equation is the integral
\be
F = a^3 \epsilon \, C_2 + 2 A(t),
\ee
where $C_2$ is a constant of integration and $A(t)$ is a function of time defined as
\be
A(t) \equiv \frac{a^3 \epsilon}{2} \int^{t} \!\!  \frac{dt}{a^3 \epsilon H} .  \label{def-A}
\ee  
In the previous expression, $A(t)$ contains the indefinite part of the integral, without the constant part already accounted in $C_2$. The integral can be solved by iterating partial integrations infinite times. We arrive to the following formal result:

\be \label{local-A}
A(t) =-\frac{\epsilon}{3} \sum^{\infty}_{n=0} \frac{e^{3N(1-n)/2}}{(n+1)!}\left[\frac{2}{3}e^{3N/2}\frac{d}{dN} \right]^n \left(\frac{e^{-3N/2}}{H^2\epsilon}\right)
\ee

where $N = \ln a(t)$ is the usual $e$-fold number. It is easy to appreciate that $A(t)$ is a function of slow-roll parameters. The first few terms of the previous expression are:
\bea
H^2 A &=& - \frac{1}{3} + \frac{1}{6} \left( 1 - \frac{4}{3} \epsilon + \frac{2}{3} \eta \right) \nn \\
&& - \frac{1}{27} \left[ \left( 1 - \frac{4}{3} \epsilon + \frac{2}{3} \eta \right) \left( 2 \epsilon+ \eta \right) - \left(  \frac{4}{3} \epsilon \eta + \frac{2}{3} \eta \xi \right) \right] \nn \\
&&  + \cdots ,
\eea
where $\eta=\frac{\dot{\epsilon}}{\epsilon H}$ and $\xi=\frac{\dot{\eta}}{\eta H}$ are the second and third slow-roll parameters. Equation (\ref{local-A}) can be resumed back to
\be
A \simeq - \frac{1}{2(3 + \eta - 2 \epsilon) H^2 } ,  \label{A-eta}
\ee
up to corrections suppressed by $\epsilon$, $\xi$ and higher slow-roll parameters (this re-summation is valid for values of $\eta$ of order 1). It would seem that for $\eta = -3+2\epsilon$ the function $A$ becomes ill defined. However, in that case one has to resume back the neglected slow-roll parameters (such as $\epsilon$ and $\xi$). 

To continue, from Eq.~(\ref{new-xi}) we now have $\xi^0 = C_3 + 2 A(t)  \dot \zeta_L$, where $C_3 = C_1 +  C_2 a^3 \epsilon \dot \zeta_L$ [notice, from Eq.~(\ref{motion-long}), that $a^3 \epsilon \dot \zeta_L$ is a constant, and therefore $C_3$ is an overall integration constant for $\xi^0$]. The choice of $C_3$, which represents a constant shift between $t$ and $\bar t$ is, at this point, arbitrary. As we shall see in Section~\ref{sec:modulation}, where the modulation of short-wavelength modes by long-wavelength modes is discussed, a convenient choice is given by $C_3 = 0$.\footnote{This choice will ensure that the modulation of short-wavelength modes by long-wavelength modes consists of an expansion in terms of the non-adiabatic pressure, which will be suppressed if the fluctuations remain adiabatic.} With this consideration, we finally arrive to the following expression for $\xi^0$:
\be \label{xi-0}
\xi^0 =  2 A(t)  \dot \zeta_L . 
\ee
Notice that in the case of attractor models, where $\dot \zeta_L = 0$, we recover the result $\xi^0=0$. Otherwise, in the case of ultra slow-roll, one has $A=1/6H^2$ and $\dot \zeta_L \propto a^3$, from where one obtains $\xi^0(t) =  \dot \zeta_L /3 H^2$. With (\ref{xi-0}), independently of the background evolution at time $t_*$, if the system enters an attractor phase, one obtains $\xi^0=0$ and the attractor solution is recovered.

\subsection{Attractor vs non-attractor backgrounds} \label{attractor-vs-non}

It will be useful to count with a criteria to distinguish attractor and non-attractor backgrounds, making connection with the time evolution of $\zeta_L$. Note that from Eq.~(\ref{motion-long}) it follows that 
\be
\dot \zeta_L = \frac{C}{a^3 \epsilon} ,
\ee
where $C$ is an integration constant. Integrating this expression we end up with
\be
\zeta_L(t) = C_0 +  C \int^t \frac{dt}{a^3 \epsilon} , \label{zeta-super-horizon}
\ee
where $C_0$ is another integration constant. Now, attractor backgrounds are those for which the second term in (\ref{zeta-super-horizon}) consists of a decaying mode (that is, for all purposes $\zeta_L(t)$ is a constant).\footnote{By recalling that Eq.~(\ref{xi-eq}) is related to Eq.~(\ref{background-bar-1}), it is not difficult to verify this definition agrees with the more standard definition whereby $\ddot \phi(t)$ is irrelevant to determine the state of the background.} This is achieved as long as the integrated denominator grows quickly within an $e$-fold  (\emph{e.g.} an exponential grow in terms of $e$-folds). To be concrete, we define the following dimensionless parameter:
\be
\gamma \equiv \frac{1}{H} \frac{d}{dt} \ln (a^3 H \epsilon) ,  \label{gamma-parameter}
\ee
which in terms of slow roll parameters reads $\gamma = 3 + \eta - \epsilon $. This quantity parametrizes the growth (in units of $e$-folds) of the denominator inside the integral of Eq.~(\ref{zeta-super-horizon}). If $\gamma$ stays almost constant, one obtains that
\bea
\zeta_L(t) &\simeq& C_0 +  C \int^N \frac{dN}{e^{\gamma N}} \\ &\simeq&  C_0 - \frac{C}{\gamma} e^{-\gamma N} .
\eea
In this way, attractor models are characterized by $\gamma  \gtrsim 1$ whereas non-attractor models are characterized by  $\gamma \lesssim -1$. For instance, in single field slow roll inflation, $\epsilon$ stays almost constant, and $a(t)$ grows exponentially, in which case $\gamma \simeq 3$. On the other hand, in ultra slow-roll  $\epsilon$ decreases as $a^{-6}$, from which one sees that $\gamma = - 3$ implying a dramatic growth of $\zeta_L(t)$. As we have already seen, regardless of the evolution of the background, $\zeta_L(t)$ always can be absorbed in the background. We exploit this fact in Section~\ref{sec:modulation}.


\section{Modulation of short wavelengths in co-moving coordinates}  \label{sec:modulation}

Let us now examine how the compelling results of the previous discussion can be used to derive both (\ref{maldacena-relation}) and (\ref{maldacena-relation-violated}) as particular examples of a single result. To start with, let us denote 
\be
\zeta(t,\x) = \zeta [t, \x ; a (t)] , \label{zeta-background-a}
\ee 
as the solution of the perturbed metric (\ref{pert-metric-original}), where the notation emphasizes that $\zeta(t,\x)$ is a solution to a perturbed system with a background scale factor $a(t)$. This definition includes a prescription to fix the initial conditions of $\zeta(t,\x)$. Now, notice that $\bar \zeta$ of (\ref{pert-metric-new}) must be a solution of the same equations of motion governing the system (\ref{pert-metric-original}) but with coordinates $(\bar t,\bar {\x})$ and a scale factor $\bar a(\bar t)$, instead of $(t,\x)$ and $a(t)$. That is 
\be
\bar \zeta (\bar t , \bar \x)  = \zeta[\bar t, \bar {\x} ; \bar a (\bar t)].
\ee 
But thanks to~(\ref{identification-zeta-S-zeta-bar}) we have 
\be
\zeta_S ( t ,  \x) = \bar \zeta ( \bar t , \bar \x) = \zeta[\bar t, \bar {\x} ; \bar a (\bar t)] , \label{zeta_S-zeta_bar}
\ee 
where it is understood that $(\bar t, \bar \x)$ appearing in $\bar \zeta$ can be expressed in terms of $(t,\x)$ due to the change of coordinates determined by $\xi^0$ and $\beta$. Therefore, it follows that
\bea
\zeta_S ( t ,  \x)  &=& \zeta [\bar t , \bar {\x} ; \bar a (\bar t)]  \nn \\
&=& \zeta [  t + \xi^0  , e^{\beta/3} \x ;  a ( t ) e^{\alpha}]  \nn \\
&=& \zeta [ t + \xi^0, e^{\beta/3} \x ;  a ( t + \xi^0 ) e^{D_L}] , \qquad\qquad \label{zeta_S-modulation-steps}
\eea
where in the last line we have defined $D_L$ as 
\be
D_L \equiv \alpha - H \xi^0 . \label{D_L-def}
\ee

Let us briefly consider the time derivative of this quantity, given by $\dot D_L = \dot \alpha - \dot H \xi^0 - H \dot \xi^0$. By employing Eqs.~(\ref{cond-perts-again2-1})-(\ref{cond-perts-again2-3}) to express $\dot \alpha$ and $\dot \xi^0$ in terms of $\zeta_L$, and using (\ref{xi-0}), one gets
\be \label{dot-D_L}
\dot D_L = \left(2H^2A(t)-\frac{1}{3}\right)\epsilon \dot \zeta_L .
\ee
By recalling the re-summed expression for $A(t)$ given in Eq.~(\ref{A-eta}), we then obtain
\be \label{D_L-pnad}
\dot D_L = \frac{\delta p_{\rm nad}}{2H(3+\eta-2\epsilon)},
\ee
where we have identified the non-adiabatic pressure $\delta p_{\rm nad}$ on large-scales as $\delta p_\text{nad} = \frac{2}{3}\left( -6 +2\epsilon -\eta \right)\epsilon  H\dot{\zeta}_L$ (see for instance~\cite{Romano:2015vxz}). In models where $\delta p_{\rm nad}=0$, $\zeta$ is called an adiabatic fluctuation, and its evolution remains unaffected by the details of short wavelength physics at the end of inflation (e.g. reheating)~\cite{Weinberg:2003sw}, ensuring that inflation is the exclusive responsible for the adiabatic initial conditions required by the Hot Big-Bang cosmology. Thus, our result~(\ref{D_L-pnad})
shows that if we restrict ourselves to situations where the fluctuations evolve adiabatically\footnote{In slow-roll inflation, this condition is automatically satisfied since $\dot{\zeta}_L=0$. During ultra slow-roll, although $\dot{\zeta}_L\neq 0$, one finds $\eta = -6+2\epsilon$ \cite{Mooij:2015yka}, and again, the non-adiabatic pressure vanishes exactly. Departures of this condition are found, for instance, during abrupt transition between SR and USR regimes.} then $D_L$ stays constant.\footnote{We note that this is true thanks to the choice $C_3 = 0$ made in Section~\ref{sec:xi0}.}  This in turn, allows us to write:
\be
\zeta[t, \x ; a(t)] = \zeta[t , \x e^{D_L}; a (t) e^{-D_L}] , \label{zeta-scaling-DL}
\ee 
which follows from the fact that in the equation of motion for $\zeta$ the quantities $a$ and $\x$ appear together through the combination $a^{-2} \partial_{\x}^2$ (therefore we can simultaneously reescale both quantities by a constant without changing $a^{-2} \partial_{\x}^2$). 

Otherwise, for models where $\delta p_{\rm nad}$ is suppressed, we expect small corrections to (\ref{zeta-scaling-DL}) of the form $
\zeta[t, \x , a(t)] = \zeta[t , \x e^{D_L}, a (t) e^{-D_L}] + \mathcal O (\delta p_{\rm nad})$. Disregarding corrections of order $\delta p_{\rm nad}$, our previous result allows us to go back to Eq.~(\ref{zeta_S-modulation-steps}) and write
\bea
\zeta_S ( t ,  \x) &=&  \zeta [ t + \xi^0, e^{\beta/3 + \alpha - H \xi^0} \x ,  a ( t + \xi^0 ) ] , \nn \\
&=&  \zeta ( t + \xi^0, e^{\beta/3 + \alpha - H \xi^0} \x ) , \nn \\
&=& \zeta (  t + 2 A  \dot \zeta_L , e^{\zeta_L -  2 A H  \dot \zeta_L } \x ) ,  \label{short2}
\eea
where in the last step we used Eqs.~(\ref{cond-perts-again2-2}) and (\ref{xi-0}). In Appendix~\ref{classicalization}, we show that $[  \zeta, \dot{\zeta}] / (\langle \zeta^2 \rangle \langle \dot{ \zeta}^2 \rangle )^{1/2} \to 0$ on superhorizon scales for any value of $\gamma$ [as defined in Eq.~(\ref{gamma-parameter})], implying that $\zeta_L$ behaves as a classical field. As a result, Eq.~(\ref{short2}) shows how $\zeta_L$ acts as a classical field that modulates $\zeta_S$. From here, it is direct to derive the following expression for the bispectrum in co-moving coordinates (see for instance~\cite{Cheung:2007sv}). To proceed, let us use the notation $\langle \zeta (t , \x) \zeta (t  , \x') \rangle  =  \langle \zeta \zeta \rangle (t, |\x - \x' |) $. Then, from (\ref{short2}) the two-point correlation function of $\zeta_S ( t , \x) $ is given by
\be
\langle \zeta_S  \zeta_S  \rangle (t , |\x - \y|) = \langle \zeta  \zeta  \rangle ( t  + 2 A  \dot \zeta_L  , e^{\zeta_L -  2 H A  \dot \zeta_L } |\x - \y|)  .
\ee
Expanding this expression, and writing it in Fourier space, we obtain
\bea
 \langle \zeta_S  \zeta_S  \rangle  (\k_1 , \k_2) =  \langle \zeta  \zeta  \rangle  (\k_1 , \k_2) + 2 A  \dot \zeta_L (\k_L) \dot P_{\zeta} (k_S , t) \qquad \nn \\
  - \left[ \zeta_L (\k_L) - 2 H A  \dot \zeta_L (\k_L)  \right]  (n_s - 1) P_{\zeta} (k_S , t) , \label{intermediate-consistency} \qquad
\eea
where $\k_S \equiv (\k_1 - \k_2)/2$ and $\k_L \equiv \k_1 + \k_2$. In the previous expression, $n_s - 1 \equiv \frac{\partial}{\partial \ln k} \ln \left[ k^3 P_{\zeta} (k , t) \right]$ is the spectral index of the power spectrum $ P_{\zeta} (k_S , t)$.  Then, correlating (\ref{intermediate-consistency}) with a long mode $\zeta_L (\k_3)$, and using $\langle \zeta_{L} (\k_3)  \langle \zeta_S  \zeta_S  \rangle  (\k_1 , \k_2)  \rangle  = \lim_{k_3 \to 0} (2 \pi)^3 \delta(\k_1 + \k_2 + \k_3) B_\zeta (\k_1 , \k_2 , \k_3)$, to identify the squeezed limit of the bispectrum, we finally obtain:
\bea
\lim_{k_3 \to 0}  B_\zeta (\k_1 , \k_2 , \k_3) &=& -  (n_s - 1) P_{\zeta} (k_L )  P_{\zeta} (k_S) \nn \\
&& \!\!\!\!\!\!\!\!\!\!\!\!\!\!\!\!\!\!\!\!\!\!\!\!\!\!\!\!\!\!\!\!\!\!\!\!\!\!\!\!\!\! +  A \dot P_{\zeta} (k_L ) \left[  \dot P_{\zeta} (k_S ) + H  (n_s - 1)  P_{\zeta} (k_S )  \right]  . \label{our-result-2}
\eea
This is one of our main result. It gives the consistency relation for canonical single field inflation valid to all orders in slow-roll parameters in co-moving coordinates. 

Noteworthily, this result displays the same form of the consistency relation found in Ref.~\cite{Finelli:2017fml}, valid for scalar field theories with an exact shift symmetry (see also~\cite{Avis:2019eav, Green:2020ebl}). There, the factor $A$ takes the particular form $A = - \dot \phi / 2 \ddot \phi \, \Theta$, where $\Theta$ is a time dependent function that generalizes the constraint equation (\ref{constraint-1}) as $\delta N_L =  \dot  \zeta_L  / \Theta$ to incorporate non-canonical theories. 

Equation~(\ref{our-result-2}) implies that non-Gaussianity can be large as long as $\dot P_\zeta (k_L)$ is sizable. In the particular case of ultra-slow roll one has $\eta = -6 +2\epsilon$, and $\zeta_L$ grows as $a^3$, implying that $\dot P_{\zeta} (k ) = 6 H P_{\zeta} (k )$. Also, thanks to (\ref{A-eta}) we see that this corresponds to the case $A = 1/H^2 6$,  from where it follows that (\ref{maldacena-relation-violated}) is a particular case of~(\ref{our-result-2}). However, as soon as the non-attractor phase finishes, and inflation goes back to a more standard attractor phase, one has $\dot P_\zeta (k_L)=0$ and the standard Maldacena's consistency relation (\ref{maldacena-relation}) is recovered. 

To finish this section, recall that Eq.~(\ref{our-result-2}) was derived assuming that $D_L$ defined in (\ref{D_L-def}) is nearly constant, reflecting the fact that $\delta p_{\rm nad}$ is suppressed for long-wavelength perturbations (and therefore, that the evolution of $\zeta_L$ is adiabatic). Of course, this is not automatically ensured by the general dynamics of $\zeta_L$. For instance, a sharp transition between ultra-slow roll and slow-roll phases can affect the evolution of $D_L$ by producing a large non-adiabatic pressure. For instance, in~\cite{Cai:2017bxr} it was found that the bispectrum can be enhanced if such transitions happen, but it becomes suppressed if the transitions are smooth, in agreement with our result~(\ref{our-result-2}). As a matter of fact, the obstruction of taking $D_L$ as a constant was addressed in~\cite{Suyama:2021adn}, and an alternative result for the bispectrum in co-moving coordinates was obtained, in agreement with that of~\cite{Cai:2017bxr}. We notice that $a(t)$ in our definition~(\ref{zeta-background-a}) is representative of the background in general, as the equation of motion for $\zeta$ will contain $a(t)$ and further time derivatives of it (\emph{i.e.} $H$, $\epsilon$ and $\eta$). In Ref.~\cite{Suyama:2021adn} a second background quantity $\dot \phi$, in addition to $a(t)$, was considered in the definition~(\ref{zeta-background-a}). With that second quantity, our step performed in Eq.~(\ref{zeta_S-modulation-steps}) involves a new background function $\frac{\partial \bar \phi}{\partial \bar t} = (1 - \dot \xi^0 - \frac{\ddot \phi}{\dot \phi} \xi^0 ) \dot \phi (t + \xi^0 ) = (1 - \dot \xi^0  ) \dot \phi (t)$ in addition to $\bar a (\bar t) = a (t + \xi^0 ) e^{D_L}$. However, as we have shown, the pair $\frac{\partial \bar \phi}{\partial \bar t}$ and $\bar a(\bar t)$ respect the same equations as $\dot \phi (t)$ and $a(t)$, thus it is enough to keep $\bar a (\bar t)$ in ~(\ref{zeta-background-a}) with the understanding that $H$, $\epsilon$ and $\eta$ are derived from $a$. It is possible to show that 
\be
\frac{\partial \bar \phi}{\partial \bar t} = \left( 1+ \frac{3}{\epsilon H}\dot D_L - \frac{6+\eta-2\epsilon}{2\epsilon H^2 (\eta -2\epsilon)} \ddot D_L \right)\dot \phi, \label{phi-DL}
\ee
from where one sees that by considering $\frac{\partial \bar \phi}{\partial \bar t}$ as an additional quantity in~(\ref{zeta-background-a}) allows one to see explicitly the role of time derivatives of $D_L$ in the modulation. However, if $\delta p_{\rm nad}$ remains suppressed, and therefore $D_L$ remains constant, one can perform the step of Eq.~(\ref{zeta-scaling-DL}) without worrying about the role $\dot \phi$.

\section{Conformal Fermi Coordinates}  \label{Sec:CFC}

We now consider the task of moving from co-moving coordinates to conformal Fermi coordinates. As explained in detail in Ref.~\cite{Pajer:2013ana} (see also~\cite{Dai:2015rda,Cabass:2016cgp}), these are coordinates that describe the local environment of inertial observers. To make this discussion easy to compare with the existing literature, here we adopt conformal time $\tau$, defined through the relation $d t = a(\tau) d \tau$, where the scale factor $a(\tau)$ (with $\tau$ as an argument) should be understood as the composition function $a(\tau) \equiv a(t (\tau))$. With this, using (\ref{identification-zeta-S-zeta-bar}), the metric (\ref{metric-step}) takes the form
\bea
ds^2 &=&  a^2 (\tau) \Bigg[ - e^{2 ( \delta \N_S + {\xi^0}' / a (\tau))} d\tau^2 \nn \\ 
&& \!\!\!\!\!\!\!\!\!\!\!\! +   e^{2 ( \zeta_S + \alpha + \beta / 3)}\left( d \x  +   a(\tau) {\bN}_S d\tau + \frac{1}{3} \x \beta'   d\tau  \right)^2 \Bigg] , \quad  \label{pert-metric-conf}
\eea
where primes ($'$) denote derivatives with respect to conformal time. To re-express this metric in CFC's we need to consider the following change of coordinates from the conformal co-moving coordinates $(\tau, \x)$ to the conformal Fermi coordinates $(\tau_F, \x_F)$
\be
\tau = \tau_F + \xi^0_F ,  \qquad  \x =  \x_F + \boldsymbol{\xi}_F , 
\ee
with~\cite{Cabass:2016cgp, Bravo:2017gct}
\bea
\xi^0_F &=& \! \int^\tau \!\!\!\! ds \left[ \frac{a_F (s)}{a(\tau (s))} -1 - \frac{1}{\mathcal H} \zeta_L' (s) \right] + \mathcal O (\x_F) , \label{CFC-1} \\
\boldsymbol{\xi}_F &=& \!  \left[ \frac{a_F (\tau_F)}{a(\tau)} -1 - \zeta_L (\tau) \right] \! \x_F   + \mathcal O (\x_F^2)  , \quad \label{CFC-2}
\eea
where we have only included the scalar contributions to $\xi^0_F$ and $\xi^i_F$.  In the previous expressions, $\mathcal H \equiv  a' / a = a H$. In addition, $\mathcal O (\x_F)$ and $\mathcal O (\x_F^2)$ stand for terms linear in $\zeta_L$ that contribute to the appearance of the projection effects in (\ref{B-obs=0}). On the other hand, the function $a_F (\tau_F)$ is the scale factor experienced by the inertial observer, given by
\be
a_F (\tau_F) = a(\tau) \left[ 1 + \zeta_L + \frac{1}{3} \int^\tau \!\!\!\! ds \, \partial_i V^i \right] ,  \label{aF}
\ee
where $V^i$ are the spatial components of the 4-velocity of the inertial observer in co-moving coordinates, given in terms of $\zeta_L$ as
\be
V^i =  -\frac{1}{3} \epsilon \zeta_L' x_F^i  + \mathcal O (\x_F^2) . \label{V}
\ee
For comparison, Eqs. (\ref{CFC-1}), (\ref{CFC-2}), (\ref{aF}) and (\ref{V})  correspond to the respective Eqs. (2.19a), (2.19b), (2.15) and (2.14) of Ref.~\cite{Cabass:2016cgp} (or Eqs. (2.34), (2.37), (2.30) and (2.8) of Ref.~\cite{Bravo:2017gct}).\footnote{A technical note for readers interested in matching our expressions with those of Refs.~\cite{Cabass:2016cgp} and~\cite{Bravo:2017gct}: Notice that both in~\cite{Cabass:2016cgp} and~\cite{Bravo:2017gct} the velocity field $V^i$ appears in the definition of $\xi^i_F$ (c.f. Eq. (2.19a) of Ref.~\cite{Cabass:2016cgp} and Eq. (2.35) in Ref.~\cite{Bravo:2017gct}). However, it is the vector part $V_v^i$ of $V^i$ that appears in those definitions (with $\partial_i V_v^i = 0$), and not the full velocity $V^i$. This is because the scalar part $V_s^i$ of $V^i$ is already accounted in (\ref{CFC-2}) as the coefficients multiplying $x_F^ i$, thanks to Eq.~(\ref{aF}). Given that here we are only interested in scalar modes, $V^i$ does not appear explicitly in Eqs. (\ref{CFC-1}) and (\ref{CFC-2}).} Now, from (\ref{cond-perts-again2-3}) notice that $\epsilon \zeta_L'$ is nothing but $\beta'$. Thus, we can integrate $ \int^\tau \!\!\!\! ds \, \partial_i V^i $ in (\ref{aF}) to obtain $a_F (\tau_F) = a(\tau) \left[ 1 + \zeta_L -   \beta /3   \right]$. Then, using (\ref{cond-perts-again2-2}), it follows that $a_F (\tau_F)$ coincides with $\bar a (\bar \tau)$:
\be
a_F (\tau_F) = \bar a (\bar \tau). 
\ee
Moreover, the diffeomorphism leading to the CFC's takes the form
\bea
\xi^0_F &=& \! \int^\tau \!\!\!\! ds \left[ \alpha - \frac{1}{\mathcal H} \zeta_L' (s) \right] + \mathcal O (\x_F) , \\
\boldsymbol{\xi}_F &=& \!  - \frac{1}{3} \beta  \x_F  + \mathcal O (\x_F^2)  . \quad
\eea
Using these results back into (\ref{pert-metric-conf}) we see that the diffeomorphisms $\xi^0_F$ and $\boldsymbol{\xi}_F$ cancel out with the various functions of $\zeta_L$ (such as ${\xi^0}'$, $\alpha$, $\beta$, and $\beta'$) and we finally obtain
\bea
ds^2 &=&  a_F^2(\tau_F) \bigg[  - e^{2 \delta  \N_S} d \tau_F^2  \nn  \\
&&  +  e^{2 \zeta_S } \left(d {\x_F} + a_F(\tau_F) {\bN}_S  d \tau_F \right)^2 \bigg]  + \cdots , \qquad \label{pert-metric-conf-new} 
\eea
where the elipses stand for terms that give rise to projection effects. This result shows that the CFC frame coincides with the frame studied in Section~\ref{sec:time-diff-perts}, in which the long wavelength perturbation is completely absorbed in the background. 

Now, it is useful to compare this form of the metric with the one obtained by directly performing the change of coordinates from co-moving coordinates to conformal Fermi coordinates. The metric components in the CFC frame can be obtained as
\be
g^{F}_{\bar \mu \bar \nu} = \frac{\partial x^{\mu}}{\partial x^{\bar \mu}}  \frac{\partial x^{\nu}}{\partial x^{\bar \nu}} g_{\mu \nu} .
\ee
One then obtains that the short- and long-wavelength contributions of the curvature perturbation $\zeta^F$ in the CFC frame are related to the short- and long-wavelength contributions of the curvature perturbation $\zeta$ in co-moving coordinates as~\cite{Cabass:2016cgp}:
\bea
 \zeta_S^{F} (\tau_F , \x_F ) &=&  \zeta_S (\tau , \x)  , \label{zeta-F_S-zeta_S}   \\
 \zeta_L^{F} (\tau_F  ) &=& \zeta_L - \frac{1}{3} \beta - H \xi^0 .   \label{zeta-F_L-zeta_L}
\eea
Notice that thanks to (\ref{cond-perts-again2-2}), the second equation is equivalent to
\be
\zeta_L^{F} (\tau_F  ) = \alpha - H \xi^0 , \label{zeta_L-alpha}
\ee 
from where it follows that
\be
a_F(\tau_F)  = a(\tau_F) e^{ \zeta^F_L} . \label{a_F-a-F}
\ee
This allows one to re-write the metric line element (\ref{pert-metric-conf-new}) as
\bea
ds^2 &=&  a^2(\tau_F) e^{2 \zeta^F_L} \bigg[  - e^{2 \delta  \N_S} d \tau_F^2  \nn  \\
&&  \qquad +  e^{2 \zeta^F_S } \left(d {\x_F} + a(\tau_F) {\bN}_S  d \tau_F \right)^2 \bigg]  + \cdots . \qquad \label{pert-metric-conf-new-2} 
\eea
The result (\ref{a_F-a-F}) shows explicitly how the background quantity $a_F(\tau_F)$ depends on long wavelength perturbations. The scale factor $a(\tau_F)$ contains no dependence on perturbations, and all long wavelength perturbations are contained in the single variable $\zeta^F_L$.

Notice that $\zeta_L^F = \alpha - H \xi^0$ is nothing but the long wavelength quantity $D_L$ defined in Section~\ref{sec:modulation}. This result shows that $\dot \zeta_L^F$ is proportional to the non-adiabatic pressure $\delta p_{\rm nad}$, from where it follows that $\zeta_L^F$ does not evolve significantly as long as the fluctuation evolve adiabatically.  This is in contrast with the evolution of $\zeta_L$ in non-attractor backgrounds, which is dominated by a growing mode. For instance, in the particular case of ultra slow-roll one has $A = 1/6 H^2$, and so $\dot \zeta_L^F = 0$.

\section{Observable primordial bispectrum's squeezed limit} \label{sec:obs-bispectrum}

Finally, we turn to the computation of the observable primordial bispectrum's squeezed limit. First, let us notice that according to (\ref{zeta-F_S-zeta_S}) the two point function of the short-wavelength modes of $\zeta$ in the CFC frame is given by
\bea
\langle \zeta^F_S \zeta^F_S   \rangle  (\tau_F, |\x_F - \x_F' |)  &=&  \langle \zeta_S \zeta_S \rangle (\tau, |\x - \x' |)  .
\eea
But recall from (\ref{zeta_S-zeta_bar}) that $\zeta_S ( t ,  \x)  = \zeta [\bar t , \bar {\x} ; \bar a (\bar t)] $. In terms of conformal Fermi coordinates, this means
\be
\zeta_S ( \tau_F ,  \x_F)  = \zeta [\tau_F , \x_F ;  a_F (\tau_F)] .  \label{zeta_S-FF}
\ee
Let us see now how this result can be used to compute the observable squeezed limit of the bispectrum. We start by examining the standard case of single field slow-roll inflation. We then move on to examine more general backgrounds.

\subsection{Attractor backgrounds}

In canonical single field inflation, attractor models are characterized for having their long wavelength perturbations $\zeta_L$ constant. From the discussion of Section~\ref{sec:initial-conditions}, this implies the following constant values for $\alpha$, $\beta$ and $\xi^0$: 
\bea
\alpha &=& 0 , \\ 
\beta &=& 3 \zeta_L , \\
 \xi^0 &=& 0.
\eea
Thus, Eq.~(\ref{zeta-F_L-zeta_L}) tells us that the long-wavelength mode $\zeta_L^F$ in the CFC frame vanishes and, thanks to (\ref{a_F-a-F}), the scale factor $a_F(\tau_F)$ of the CFC frame acquires no dependence on long wavelength perturbations. Using the notation of Section~\ref{sec:modulation}, we can write
\be
\zeta_S ( \tau_F ,  \x_F)  = \zeta [\tau_F , \x_F ,  a (\tau_F)] = \zeta (\tau_F , \x_F ) ,
\ee
which shows that the solution $\zeta_S ( \tau_F ,  \x_F) $ is simply the solution $\zeta (\tau , \x)$ with the coordinates $(\tau , \x)$ replaced by $(\tau_F , \x_F)$, unaffected by the long wavelength perturbation $\zeta_L$. As a result, the two point function of $\zeta^F_S$ is simply given as
\bea
\langle \zeta^F_S \zeta^F_S   \rangle  (\tau_F, |\x_F - \x_F' |)  &=&  \langle \zeta \zeta \rangle (\tau_F, |\x_F - \x'_F |)  .
\eea
We therefore re-obtain the well known result that in single field inflation the two point function $\langle \zeta^F_S \zeta^F_S   \rangle  (\tau_F, |\x_F - \x_F' |)$ is not modulated by $\zeta_L$. This, in turn, leads to (\ref{B-obs=0}).

\subsection{Non-attractor backgrounds} \label{sec:non-attractor-backgrounds}

Because $\zeta^F_L = D_L = \alpha - H \xi^0$, if the evolution of $\zeta_L$ is adiabatic (that is, if $\delta p_{\rm nad}$ remains supressed), then we can write $\zeta^F_L \simeq - 2 H_* A_* \dot \zeta^*_L$. Using the fact that $\epsilon a^3 \dot \zeta$ is a constant, then:
\be
\zeta^F_L = -  \frac{2 H_* A_*}{\epsilon_* a_*^3} \epsilon a^3 \dot \zeta_L . \label{zeta-F_L-*}
\ee
On the other hand, Eq.~(\ref{zeta-super-horizon}) with (\ref{gamma-parameter}) allows us to write $\zeta_L$ as
\be
\zeta_L (t) = \zeta_L(t_*) + \frac{\epsilon a^3 \dot \zeta_L}{\epsilon_* a_*^3 H_* } \int^N_{N_*} dN' \exp \left[- \int^{N'}_{N_*} \!\! \gamma \, dN \right] . \label{zeta_L-*}
\ee
Here, as discussed in Section~\ref{attractor-vs-non} the second term is a decaying solution if the background is attractor, or a growing solution in the case the background is non-attractor. Thus, putting together Eqs.~(\ref{zeta-F_L-*}) and (\ref{zeta_L-*}), we obtain the following general expression relating $\zeta_L^F (t)$ and $ \zeta_L (t)$
\be
 \zeta_L^F (t) =   \left[ \zeta_L(t) - \zeta_L(t_*) \right]  G(t,t_*) , 
\ee
where
\be
 G(t,t_*) \equiv-  \frac{2 H_*^2 A_*}{ \int^N_{N_*} dN' \exp \left[- \int^{N'}_{N_*} \!\! \gamma \, dN \right] }  .
\ee
If inflation experiences a long enough period of non-attractor evolution, we expect $\zeta_L$ to become dominated by the growing solution. Thus we can simply write:
\be
 \zeta_L^F (t) \simeq  \zeta_L(t) G(t,t_*) .  \label{zeta_L-G}
\ee 
Now, given that $\zeta_L^F (t)$ stays almost constant, the rapid growth of $ \zeta_L(t)$ during a non-attractor phase is characterized by the fact that $G(t,t_*)$ dilutes quickly within a couple of $e$-folds:
\be
 G(t,t_*) \to 0 . \label{Bto0}
\ee
For instance, if $\gamma$ stays almost constant, then for $(N-N_*) \gamma \gg 1$ one finds
\be
 G(t,t_*) \simeq  2 H_*^2 A_* \gamma  e^{ (N-N_*)  \gamma } 
 ,
\ee
which for non-attractor models satisfies (\ref{Bto0}). To continue, let us first assume that the non-addiabatic pressure does not experiences sudden changes. In this case, we can treat $\zeta^F_L $ to be nearly constant. Then,  starting with  (\ref{zeta_S-FF}) we can write 
\bea
\zeta_S ( \tau_F ,  \x_F)  &=& \zeta [\tau_F , \x_F ,  a (\tau_F) e^{\zeta^F_L}  ] , \nn  \\
&\simeq& \zeta [\tau_F , \x_F e^{\zeta^F_L} ,  a (\tau_F)   ] , \nn  \\
&\simeq& \zeta (\tau_F , \x_F e^{\zeta^F_L} ) , \label{zeta_S-steps}
\eea
where, just as in we did in Section~\ref{sec:modulation}, we used the fact that $\zeta[t, \x , a(t)] = \zeta[t , \x e^{D}, a (t) e^{-D}]$ for a constant $D$ (we shall soon address the size of the corrections implied by the fact that $\zeta_L^F$ does not stay exactly constant). It therefore follows that the two point function $\langle \zeta^F_S \zeta^F_S   \rangle $ is modulated by $\zeta_L^F$. Expanding it to first order in $\zeta_L^F$ we obtain:
\bea
&& \langle \zeta^F_S \zeta^F_S   \rangle  (\tau_F, r_F)  =  \langle \zeta \zeta \rangle (\tau_F,  r_F e^{\zeta^F_L}  ) \nn  \\
&& \qquad =  \langle \zeta \zeta \rangle (\tau_F,  r_F   ) + \zeta^F_L \frac{\partial}{\partial r_F } \langle \zeta \zeta \rangle (\tau_F, r_F  )  + \cdots . \quad \label{two-point-zeta_S-F}
\eea
Now, in~\cite{Pajer:2013ana} (see also~\cite{Dai:2015rda}) it is argued that the squeezed limit of the bispectrum in the CFC frame may be obtained by correlating $\langle \zeta^F_S \zeta^F_S   \rangle $ with $\zeta_L$ (not $\zeta_L^F$). That is, the observable squeezed limit is related to the correlation between $\zeta_L$ and $\langle \zeta^F_S \zeta^F_S   \rangle $ as:
\bea
\langle \zeta_L (\k_3) \langle \zeta_S^F (\k_3) \zeta_S^F (\k_3) \rangle  \rangle \equiv \qquad \qquad \qquad \qquad  \nn \\
 (2 \pi)^3 \delta (\k_1 + \k_2 + \k_3) B_\zeta^{\rm obs} (\k_1 , \k_2 , \k_3) . \label{def-bispectrum-CFC-1}
\eea
Inserting~(\ref{two-point-zeta_S-F}) in (\ref{def-bispectrum-CFC-1}) and using our previous result (\ref{zeta_L-G}) we are then finally led to 
\bea
\lim_{k_3 \to 0}  B^{\rm obs}_\zeta (\k_1 , \k_2 , \k_3)   \qquad \qquad \qquad \qquad \qquad \quad \nn \\
= -  (n_s - 1)  P_\zeta  (k_L )  P_{\zeta} (k_S)    G(t,t_*)  .  \label{Bispectrum-B-suppressed-1}
\eea
Thus, thanks to the fact that $G$ decays quickly during a non-attractor phase, we finally re-obtain (\ref{B-obs=0}). This result shows that non-attractor phases do not, per se, imply large local non-Gaussianity (a point already stressed in~\cite{Cai:2017bxr}). However, even though $\zeta_L^F$ remains almost constant during the whole period of inflation (as long as $\epsilon$ stays small), its evolution does not need to be smooth, implying that its role as a background quantity for the evolution of $\zeta_S^F$ may have a significant impact in the generation of local non-Gaussianity.

\subsection{Large observable non-Gaussianity?} \label{sec:LargeNG}

Let us now consider those situations in which the steps followed in~(\ref{zeta_S-steps}) cannot be performed. As we have discussed in Section~\ref{sec:modulation}, this step requires $\zeta_L^F$ to stay almost constant, which is equivalent to have a small non-adiabatic pressure. We now show that if derivatives of $\zeta_L^F$ cannot be neglected, then $B^{\rm obs}_\zeta$ can acquire large contributions (for instance, as it happens with sudden transitions). To analyze this case we may resort to the in-in formalism: We must use the metric~(\ref{pert-metric-conf-new-2}) to obtain the full action $S$ of $\zeta_S^F$. In this action, $\zeta_L^F$ appears as a background quantity. We can then identify the cubic term in $S$ containing $\zeta_S^F$ at quadratic order, and $\zeta_L^F$ at linear order. This gives the non-linear interaction sourcing the squeezed limit of the bispectrum. The action to consider is just the free field action for $\zeta_S^F$ written in the CFC frame:
\be
S = \int d^4 x_F a_F^2 (\tau_F) \epsilon_F (\tau_F) \left[ ({\zeta^F_S}')^2 - (\nabla \zeta^F_S )^2 \right]  + \cdots.
\ee
Here $a_F (\tau_F) = a (\tau_F) e^{\zeta^F_L}$ and $\epsilon_F(\tau_F)$ is computed from $a_F (\tau_F)$. One finds 
\be
\epsilon_F(\tau_F) = \epsilon (\tau_F) (1 - 2 \dot \zeta^F_L/H) - \ddot \zeta_L^F / H^2.
\ee 
This implies that the part of the action quadratic in $\zeta_S^F$ and linear in $\zeta_L^F$ is given by
\be
S_{\rm int} = \int d^4 x_F a^2 (\tau_F) \epsilon (\tau_F) \Delta_L \left[ ({\zeta^F_S}')^2 - (\nabla \zeta^F_S )^2 \right] + \cdots , \label{interaction-action-L-S}
\ee
where
\be
 \Delta_L \equiv 2 \zeta_L^F - \frac{2}{H} \dot \zeta_L^F - \frac{1}{\epsilon H^2} \ddot \zeta_L^F . \label{Delta_L}
\ee
We can now isolate the relevant part by focusing on those terms that would have the largest impact on the squeezed limit of the bispectrum given a time varying $\zeta_L^F$. This will necessarily come from the third term in (\ref{Delta_L}), which is not suppressed by $\epsilon$. To proceed, let us assume that $\epsilon$ stays small throughout the full period of inflation, and that $\eta$ is most of order $1$ (so as to keep $\epsilon$ small). Then, the largest contribution to the computation of the bispectrum comes from 
\be
S_{\rm int} \supset \int d^4 x_F a^2  \epsilon \, \eta'  \zeta_L'  ({\zeta^F_S})^2 , \label{S-int-relevant}
\ee
which is obtained from (\ref{interaction-action-L-S}) after performing partial integrations. This term is precisely what gives rise to large local non-Gaussianity in co-moving coordinates (with $\zeta_S$ instead of $\zeta_S^F$) provided that the background transit from a non-attractor phase to an attractor phase abruptly, as studied in~\cite{Cai:2017bxr}. In other words, with abrupt transitions the observed primordial bispectrum can be as large as the bispectrum computed in the co-moving frame:
\be
\lim_{k_3 \to 0}  B^{\rm obs}_\zeta (\k_1 , \k_2 , \k_3) \simeq \lim_{k_3 \to 0}  B^{\rm com}_\zeta (\k_1 , \k_2 , \k_3) . \label{obs-com-1}
\ee
For instance, as shown in~\cite{Cai:2017bxr}, in the particular case where $\eta'$ is negligible for all times except for a small period of time, the term in Eq.~(\ref{S-int-relevant}) leads to 
\be
\lim_{k_3 \to 0}  B^{\rm obs}_\zeta (\k_1 , \k_2 , \k_3) \simeq  P_\zeta  (k_3)  P_{\zeta} (k_2) \int \!\! d\tau \, \eta' ,
\ee
which, in the case of a sudden transition from ultra slow roll ($\eta \simeq - 6$) to slow roll ($|\eta| \ll 1$) gives back (\ref{maldacena-relation-violated}). 

To summarize, as we saw in Section~\ref{sec:non-attractor-backgrounds}, non-attractor backgrounds do not automatically imply a large squeezed limit of the bispectrum. The derivation of Section~\ref{sec:non-attractor-backgrounds} assumed that $\zeta_L^F$ stays almost constant, which kept open the question as to what would be the effect of having strong time variations of $\zeta_L^F$. Here we have seen that a large and sudden variation of $\zeta_L^F$ (related to a strong non-adiabatic evolution of $\zeta$) can indeed lead to a large value for the bispectrum.


\section{Discussion and conclusions}  \label{sec:conslusions}

We have analyzed the squeezed limit of the non-Gaussian bispectrum in canonical single field inflation using tools that allow certain statements to be valid at all orders in slow roll parameters. In particular, we have derived a consistency relation for the squeezed limit [c.f. Eq.~(\ref{our-result-2})] in co-moving coordinates for both, attractor and non-attractor backgrounds, which relies on the fact that the non-adiabatic pressure $\delta p_{\rm nad}$ remains suppressed and without sudden changes (in other words, that it stays small and slowly evolving), a necessary condition that can be realized for both, attractor and non-attractor models. Our approach, is limited to situations where this condition is satisfied, but subsequent work inspired by this article have studied how to apply our method to situations where $\delta p_{\rm nad}$ experiences sudden variations [see our discussion around Eq.~(\ref{phi-DL})]. We have also analyzed the computation of the bispectrum's squeezed limit in the conformal Fermi coordinate frame, which gives access to the observable squeezed limit. In general, both attractor and non-attractor backgrounds leads to a suppressed amount of local non-Gaussianity, as long as the non-adiabatic pressure remains suppressed. As we saw, large local non-Gaussianity is not a direct consequence of being in a non-attractor background. Instead, it is due to the fact that the non-adiabatic pressure experiences sudden variations, as is the case with models where $\eta'$ is momentarily large.

Our approach highlighted the important role of time diffeomorphisms in order to understand single field inflation in more general terms. As we saw, it is possible to organize perturbation theory in such a way that a time diffeomorphism does not take us away from co-moving gauge, allowing us to evade a well known obstruction of using time-diffeomorphism to study the consequences of residual diffeomorphisms~\cite{Hinterbichler:2012nm, Assassi:2012zq, Creminelli:2012qr, Hinterbichler:2013dpa}. This diffeomorphism coincides with the change of variables from co-moving coordinates to conformal Fermi coordinates, which allows the computation of observable $n$-point correlation functions. 

To finish, let us notice that in order to compute the observable primordial bispectrum we have used the approach followed in~\cite{Pajer:2013ana} whereby the observable squeezed limit of the bispectrum corresponds to (\ref{def-bispectrum-CFC-1}). However, we may alternatively consider the computation of the three point function involving only perturbations within the CFC frame:
\bea
\langle \zeta_L^F (\k_3) \langle \zeta_S^F (\k_3) \zeta_S^F (\k_3) \rangle  \rangle \equiv \qquad \qquad \qquad \qquad  \nn \\
 (2 \pi)^3 \delta (\k_1 + \k_2 + \k_3) B_\zeta^{\rm obs} (\k_1 , \k_2 , \k_3) . \label{def-bispectrum-CFC-2}
\eea
In this case, instead of obtaining (\ref{obs-com-1}) we would obtain 
\be
 B^{\rm obs}_\zeta (\k_1 , \k_2 , \k_3) \simeq  G(t,t_*)  B^{\rm com}_\zeta (\k_1 , \k_2 , \k_3)  , \label{obs-com-2}
\ee
and so, the bispectrum's squeezed limit is found to be suppressed with respect to the value computed in co-moving gauge.


\begin{acknowledgments}

We wish to thank Vicente Atal, Xingang Chen, Jorge Nore\~na, Enrico Pajer, Subodh Patil, Basti\'an Pradenas, Toni Riotto, Teruaki Suyama, Spyros Sypsas, Yuichiro Tada, Dong-Gang Wang and Masahide Yamaguchi for useful discussions and comments that helped us to improve the first version of this work. RB and GAP acknowledge support from the Fondecyt Regular Project No.~1210876 (ANID). RB acknowledges previous support from the CONICYT-PCHA Doctorado Nacional scholarship 2016-21161504 and from the FCS Swiss Government Excellence Scholarship No.~2021.0519.

\end{acknowledgments}

\appendix
\section{Quantum-to-classical transition of $\zeta_L$} \label{classicalization}
In this appendix we will verify the classical behavior of the perturbations in the long-wavelength limit, beyond the slow-roll regime.

Let us begin by considering the Mukhanov-Sasaki equation of motion for the curvature perturbation $\zeta$. With the inclusion of the parameter $\gamma$ in (\ref{gamma-parameter}), we found that it is given by 
\be \label{MS-eq}
\ddot{\zeta} +\gamma H \dot{\zeta} +\frac{k^2}{a^2}\zeta=0,
\ee
where we are assuming $\gamma \simeq \text{constant}.$
In conformal time, $d\tau=a\,dt$, and after considering the de Sitter limit $aH=-1/\tau$, $(\tau<0)$ (\ref{MS-eq}) reads 
\be\label{dS-MS}
\zeta''+\frac{(1-\gamma)}{\tau}\zeta'+k^2\zeta=0.
\ee
Introducing the change of variable 
\be\label{cov}
u = (-\tau)^{\frac{1}{2}(1-\gamma)}\zeta,
\ee
Eq.~(\ref{dS-MS}) adopts the form
\be \label{dS-nu}
u'' +\left[k^2 -\frac{1}{\tau^2}\left(\nu^2-\frac{1}{4}\right)\right]u=0,
\ee
with $\nu^2= \gamma^2/4$. The solutions of (\ref{dS-nu}), for $\nu \in \mathbb{R}$, are given by 
\be \label{MS-sol}
u_{k}(\tau)= \sqrt{-\tau}\left(C_1(k)H^{(1)}_\nu(-k\tau)+C_2(k)H^{(2)}_\nu(-k\tau)\right).
\ee
It is worth mentioning that the solution (\ref{MS-sol}) is insensitive to the sign of $\gamma$, {i.e.,} to attractor or non-attractor stages of inflation, not so $\zeta$. Now, if we impose Bunch-Davies initial conditions deep inside the horizon, 
\be
\lim_{-k\tau\gg 1}u_{k}(\tau) = \frac{e^{-ik\tau}}{\sqrt{2k}},
\ee
and considering the asymptotic behavior of Hankel functions
\begin{align}
H^{(1)}_\nu(x \gg 1)\approx \sqrt{\frac{2}{\pi x}}\,e^{i(x-\frac{\pi}{2}\nu-\frac{\pi}{4})},\\
H^{(2)}_\nu(x \gg 1)\approx \sqrt{\frac{2}{\pi x}}\,e^{-i(x-\frac{\pi}{2}\nu-\frac{\pi}{4})},
\end{align}
we can fix the integration constants as
\be \label{BD-fix}
C_2(k)=0, \quad C_1(k)=\frac{\sqrt{\pi}}{2}\,e^{i(\nu+\frac{1}{2})\frac{\pi}{2}},
\ee
thus, the final solution is
\be
u_{k}(\tau)= \frac{\sqrt{\pi}}{2}\,e^{i(\nu+\frac{1}{2})\frac{\pi}{2}}\sqrt{-\tau} H^{(1)}_\nu(-k\tau).
\ee

To continue,  we are interested in studying the quantum behavior of $\zeta$ for different evolution regimes of the background evolution. In order to do so, it is more convenient to apply the quantization procedure to the canonical variable $u$ and its conjugate momentum $\pi = u'$, requiring they satisfy the usual commutation relations,
\begin{align}
	&[\hat{u}_{\bf k}(\tau), \hat{u}_{\bf k'}(\tau)]=0, \\
	&[\hat{\pi}_{\bf k}(\tau), \hat{\pi}_{\bf k'}(\tau)]=0, \\
	&[\hat{u}_{\bf k}(\tau), \hat{\pi}_{\bf k'}(\tau)] = i (2\pi)^3 \delta({\bf k}+{\bf k'}). \label{Comm-1}
\end{align}
If we consider the mode expansion in terms of time-independent bosonic operators of the form 
\begin{gather}
    \hat{u}_{\bf k}(\tau) = u_{k}(\tau)\hat{a}_{\bf k} + u^*_{k}(\tau)\hat{a}^{\dagger}_{-\bf k}, \\
    \hat{\pi}_{\bf k}(\tau) = u'_{k}(\tau)\hat{a}_{\bf k} + u'^*_{k}(\tau)\hat{a}^{\dagger}_{-\bf k},
\end{gather}
the commutator (\ref{Comm-1}) implies 
\be
[\hat{a}_{\bf k},\hat{a}^{\dagger}_{\bf k'}] = (2\pi)^3 \delta({\bf k}+{\bf k'}),
\ee
as well as the following Wronskian normalization for the mode functions
\begin{align}
u_k u'^*_{k'}-u^{*}_k u'_{k'} = i. 
\end{align}
Let us employ the full solution (\ref{MS-sol}), considering that 
\begin{align}
u'_{k}(\tau)= &\frac{1}{\tau}\left(\frac{1}{2}+k\nu\right)u_{k}(\tau)\\&+k\sqrt{-\tau}\left(C_1(k)H^{(1)}_{\nu+1}(-k\tau)+C_2(k)H^{(2)}_{\nu+1}(-k\tau)\right),\nonumber
\end{align}
and using the following properties of Hankel functions
\begin{align}
    (H^{(1)}_{\nu}(x))^* = H^{(2)}_{\nu}(x), \quad x \in \mathbb{R}, \\
    H^{(1)}_{\nu+1}(x)H^{(2)}_{\nu}(x) - H^{(1)}_{\nu}(x)H^{(2)}_{\nu+1}(x) = - \frac{4 i}{\pi x}.
\end{align}
The Wronskian normalization requires
\be
|C_1|^2-|C_2|^2 = \frac{\pi}{4},
\ee
which is in agreement with the Bunch-Davies initial conditions (\ref{BD-fix}). 

Returning to $\zeta$, we are interested in verifying its quantum-to-classical transition on super-Hubble scales for different phases of inflation. This can be checked by computing the super-horizon behavior of the quantity
\be\label{norm-comm}
\frac{[\hat{\zeta}_{\bf k},\hat{\zeta}'_{\bf k'}]}{\sqrt{\langle \hat{\zeta}^2_{k} \rangle\langle \hat{\zeta}'^2_{k} \rangle}}. 
\ee
Since the Wronskian for the mode functions is fixed for any conformal time $\tau$, the super-horizon evolution of the commutator, will be given by (\ref{cov}). Therefore the commutator will scale like
\be
[\hat{\zeta}_{\bf k},\hat{\zeta}'_{\bf k'}] \propto (-k\tau)^{\gamma-1}.
\ee
On the other hand, as we mentioned, $u_k$ is insensitive to the sign of $\gamma$, not so $\zeta$. Therefore, considering that the Hankel function for a small argument scales like
\be
H^{(1)}_\nu(x \ll 1)\approx -i \frac{\Gamma(\nu)}{\pi}\left(\frac{2}{x}\right)^{\nu}, \quad \nu > 0, 
\ee
the super-horizon evolution of $\zeta$ will be given by
\be
\zeta_L \propto \begin{cases}
\text{const.} \quad &\gamma > 0,\\
(-k\tau)^{\gamma} \quad &\gamma < 0.\\
\end{cases}
\ee
The next step is to compute the variance of the fields, where we find
\be
\langle \hat{\zeta}^2_{k} \rangle = \langle 0 | \hat{\zeta}^2_{k}|0\rangle = (-k\tau)^{\gamma-1} |u_k|^2,
\ee
\bea
\langle \hat{\zeta}'^2_{k} \rangle &= \langle 0 | \hat{\zeta}'^2_{k}|0\rangle =(-k\tau)^{\gamma-1} \bigg[\left(\frac{\gamma-1}{2\tau}\right)^2 |u_k|^2 \nn \\
&+\frac{\gamma-1}{2\tau}(u_k u'^{*}_k+u'_k u^{*}_k) + |u'_k|^2 \bigg].
\eea
Since, we are interested in the long-wavelength limit, where
\begin{gather}
u_k \propto (-k\tau)^{1/2-\nu},\\
u'_k \propto (-k\tau)^{-1/2-\nu}, \quad \nu = |\gamma|/2.
\end{gather}
Therefore, we find 
\begin{gather}
\langle \hat{\zeta}^2_{L} \rangle \propto (-k\tau)^{\gamma-|\gamma|}, \\
\langle \hat{\zeta}'^2_{L} \rangle \propto (-k\tau)^{-2+\gamma-|\gamma|},
\end{gather}
so that, the long-wavelength behavior of (\ref{norm-comm}) scales like
\be
\frac{[\hat{\zeta}_{L},\hat{\zeta}'_{L}]}{\sqrt{\langle \hat{\zeta}^2_{L} \rangle\langle \hat{\zeta}'^2_{L} \rangle}} \rightarrow (-k\tau)^{|\gamma|}.
\ee

Then, this quantity always vanishes during the cosmic evolution whether the background is attractor or not, under the assumption of $\gamma \simeq \text{constant}$. Therefore, we can consider $\zeta_L$ as a classical field, in the sense of \cite{Assassi:2012et}.


\end{document}